\documentclass[twocolumn,preprintnumbers,amsmath,amssymb,superscriptaddress]{revtex4}

\usepackage{graphicx}
\usepackage{dcolumn}
\usepackage{bm}

\usepackage{float}
\usepackage{color}

\begin{document}

\preprint{APS/123-QED}

\title{How Graphene-like is Epitaxial Graphene? \\Quantum Oscillations and Quantum Hall Effect}
\author{Johannes Jobst}
\author{Daniel Waldmann}

\affiliation{Lehrstuhl f\"ur Angewandte Physik, Universit\"at
Erlangen-N\"urnberg, 91056 Erlangen, Germany
}%
\author{Florian Speck}
\author{Roland Hirner}

\affiliation{Lehrstuhl f\"ur Technische Physik, Universit\"at
Erlangen-N\"urnberg, 91056 Erlangen, Germany }%

\author{Duncan K. Maude}
\affiliation{Laboratoire des Champs Magn\'etiques Intenses, 25
Avenue des Martyrs, 38042 Grenoble,France}%
\author{Thomas Seyller}
\affiliation{Lehrstuhl f\"ur Technische Physik, Universit\"at
Erlangen-N\"urnberg, 91056 Erlangen, Germany }%
\author{Heiko B. Weber}
 \email{Heiko.Weber@physik.uni-erlangen.de}
\affiliation{Lehrstuhl f\"ur Angewandte Physik, Universit\"at
Erlangen-N\"urnberg, 91056 Erlangen, Germany
}%

\homepage{http://www.lap.physik.uni-erlangen.de/}
\date{\today}

\begin{abstract}
We investigate the transport properties of high-quality
single-layer graphene, epitaxially grown on a 6H-SiC(0001)
substrate. We have measured transport properties, in particular
charge carrier density, mobility, conductivity and
magnetoconductance of large samples as well as submicrometer-sized
Hall bars which are entirely lying on atomically flat substrate
terraces. The results display high mobilities, independent of
sample size and a Shubnikov-de Haas effect with a Landau level
spectrum of single-layer graphene. When gated close to the Dirac
point, the mobility increases substantially, and the graphene-like
quantum Hall effect occurs. This proves that epitaxial graphene is
ruled by the same pseudo-relativistic physics observed previously
in exfoliated graphene.

\end{abstract}

\maketitle

Graphene, a single sheet of graphite, is one of the most exciting
electronic materials in the last years \cite{rise}. The
observation of very fast charge carriers even at room temperatures
and unconventional quantum mechanics have stimulated far reaching
visions in science and technology. Many of these properties are a
direct consequence of the unique symmetry of graphene and its true
two-dimensionality. A calculation of the single-particle band
structure delivers a linear $E(k)$ dispersion relation and  a
chiral degree of freedom in the electronic wave function. This
\emph{graphene physics} modifies, for example, the quantum Hall
effect (QHE) \cite{QHE_novoselov,QHE_stormer} and efficiently
suppresses backscattering of charge carriers
\cite{Ando,Klein_tunneling_2006,Klein_tunneling_2009}.

There are today two main preparation strategies for graphene,
resulting in different materials. \emph{Mechanical exfoliation} of
single graphene sheets from graphite yields small flakes a few
tens of microns in size which are usually deposited on a silicon
wafer covered by a layer of silicon oxide allowing for
electrostatic gating. Since its discovery in 2004,
\emph{exfoliated graphene} has been the driving force for the
exploration of graphene physics. Remarkably, a broad agreement
between experiment and theory has been observed.

The second strategy is \emph{epitaxial growth} of graphene on well
defined surfaces. This procedure promises large-scale fabrication,
detailed surface-science control, and would offer technological
perspectives. Epitaxial graphene is currently developed into two
major directions. Chemical vapor deposition on Ni, for instance,
has been demonstrated to lead to graphene flakes which could be
transferred to an insulating substrate. In this case, QHE typical
for graphene was observed \cite{Kim_transfer}. Another method uses
the temperature-induced decomposition of the wide-band gap
semiconductor silicon carbide (SiC)
\cite{Forbeaux,ClaireBerger05262006,emtsev}. Since SiC can be
obtained in an insulating state, this technique does not require
transferring the graphene layer onto another substrate, which is a
clear technological advantage.

Epitaxial growth on SiC has been carried out on both polar surface
orientations. Not unexpected, the growth and the resulting layers
are dissimilar in many aspects. On the carbon terminated surface
(C-face) the decomposition is rapid and often multilayers are
grown. Transport studies of multilayered epitaxial graphene (MLEG)
on the C-face of SiC have shown SdH-oscillations of graphene
monolayers and high electron mobility, but no QHE
\cite{ClaireBerger05262006}, which is a consequence of the mutual
rotation of the graphene layers within the stack
\cite{hass,shallcross}. On the silicon-terminated face (Si-face),
growth of graphene is slower allowing for a controlled
single-layer growth. The better thickness control achieved on
Si-face SiC substrates yielding single layers is of particular
importance for top-gated field effect devices as compared to thick
stacks of MLEG \cite{Kedziersky} due to screening.

As a consequence, we have concentrated on the growth of single
layer graphene on Si-face SiC. We studied extensively its surface
with angle resolved photoemission (ARPES), scanning tunneling
microscopy, low energy electron diffraction, Raman spectroscopy,
and first transport experiments \cite{emtsev}. Altogether a
picture has been developed that this material has excellent
quality and fits well to the graphene-model band structure. An
anomaly in the ARPES spectra of epitaxial graphene on SiC(0001)
has been interpreted as the signature of many-particle
interactions \cite{Bostwick}. However, a different interpretation
of ARPES \cite{lanzara} results suggest a symmetry breaking
between A and B sublattices of graphene and subsequent formation
of a band gap induced by the presence of the so-called buffer
layer or $(6\sqrt3\times6\sqrt3)R30^\circ$ reconstruction. The
latter forms the intrinsic interface between SiC(0001) and
thermally grown graphene. The interface layer is semi-conducting,
i.e. has no states at the Fermi level, and we have proposed that
it consists of a covalently bound graphene layer
\cite{Interaction}. The higher order commensurate unit cell
(supercell) of the combined system graphene on buffer layer on
SiC(0001) contains 13$\times$13 unit cells of graphene. It remains
an unresolved question, whether this lowered symmetry spoils the
graphene physics by, e.g. inducing a band gap.

To discover the graphene physics in our system, we carried out classical Hall effect, Shubnikov-de Haas (SdH) effect and QHE measurements. The latter phenomena give a fingerprint of single-layer graphene behavior \cite{bilayer_QHE}, clearly different from parabolic band structures or non-chiral wave functions. We investigated the raw material as well as epitaxial graphene driven close to charge neutrality by chemical gating.

\begin{figure}
 \includegraphics[width=3in]{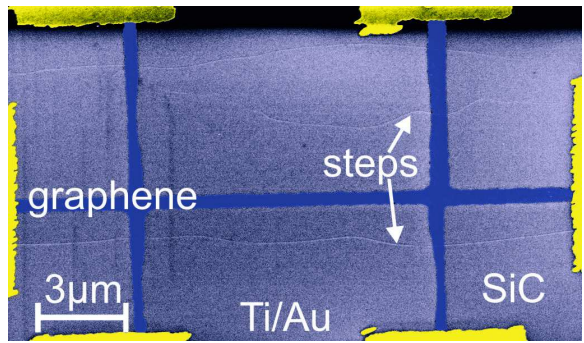}
  \caption{\label{hall_bar} Scanning electron micrograph of a Hall bar (0.48 $\mu$m width) lithographically patterned out
  of a single layer of graphene (dark blue) on an atomically flat substrate terrace of semi-insulating SiC. The substrate steps are clearly resolved.}
 \end{figure}

The growth process and the patterning has been reported in
\cite{emtsev}. Briefly, we have produced graphene on the
silicon-terminated side of semi-insulating SiC by thermal
decomposition at $1650^\circ$C and 900\,mbar argon atmosphere for
$\approx 15$\,min. Then we have removed the graphene partly by
electron beam lithography and subsequent oxygen plasma etching,
such that Hall bars of different sizes were patterned.
Fig.~\ref{hall_bar} shows a Scanning electron micrograph of a
sample with the Hall bar entirely placed on a single, atomically
flat substrate terrace of the SiC(0001) surface. Here, we obtain
reliably single-sheet graphene, as we keep some distance from the
step edges \cite{emtsev}. Other samples were much larger, included
many substrate steps and some even visible defects.

The electrical contacts were guided away from the Hall bar by
graphene leads, and further out with metallic top contacts
(Ti/Au). The samples were investigated in a cryostat fitted with a
0.66\,T magnet, or in the high-field laboratory in Grenoble in
magnetic fields up to 28\,T.

\begin{figure}
 \includegraphics[width=3in]{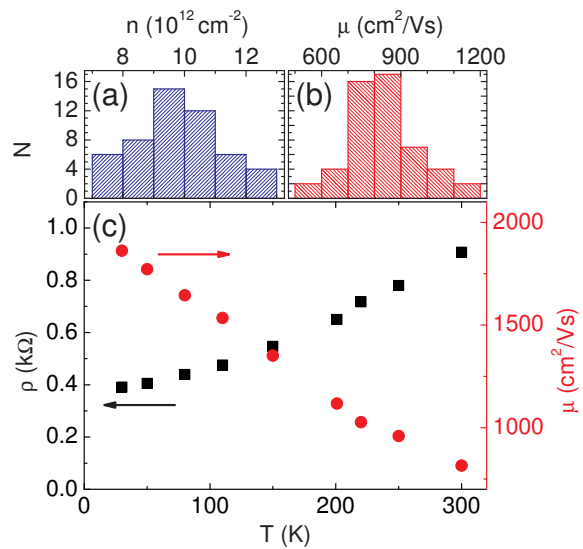}
 \caption{\label{mu_vs_n}a),b) Histograms of charge carrier density $n$ and charge carrier mobility
 $\mu$ in epitaxial graphene for 51 samples of various sizes at room temperature.
 Both quantities have been determined by Hall effect measurements.
 c)Temperature dependence of resistivity $\rho$ and mobility $\mu$ of a typical Hall-bar sample.
 The resistivity increases super-linearly with increasing $T$, while $\mu$ decreases linearly}
 \end{figure}

The quantities which can be derived from Hall bar measurements are
charge carrier densities and charge carrier mobilities.
Figure~\ref{mu_vs_n} shows data derived for 51 samples of
different sizes with the raw material. The charge carrier density
of $n \approx 10^{13}$\,cm$^{-2}$ and a mobility at room
temperature of $\mu \approx 900$\,cm$^2$/Vs is found for all
samples. The charge carrier density can be related to
electron-like transport with a chemical potential $\approx
380$\,meV above the Dirac point. This value is slightly below the
photoemission result of $\approx450$\,meV \cite{emtsev}.
Surprisingly, the mobility of rather large samples and of Hall
bars placed on a single substrate step does not differ noticeably.
Although it is known that graphene grows over step edges without
being disrupted \cite{lauffer,ClaireBerger05262006}, it is
surprising that the inhomogeneity does not affect global transport
properties significantly.

Graphene is a very surface sensitive material. Hence, one may
believe that the limitation of mobility might be caused by surface
adsorbates. We heated up four samples to 350$^\circ$C  for 30\,min
in cryogenic vacuum, in order to desorb potential adsorbates and
continued the measurement without breaking the cryogenic vacuum.
The measured quantities $n$ and $\mu$ remained essentially
unaltered. Hence, adsorbates do not play a major role in our
experiments. Further insight is gained from the temperature
dependence of resistivity. It shows a super-linear behavior as
reported for exfoliated graphene \cite{intrinsic_mobility}, but
the temperature-dependent contribution
$\Delta\rho=\rho$(300\,K$)-\rho$(0\,K$)\approx 500\,\Omega$ is one
order of magnitude larger. Hence, even if the residual resistivity
$\rho$(0\,K) could be eliminated by improved sample preparation,
the temperature dependent scattering mechanism would limit the
room temperature mobility to $\mu(300$\,K$)\approx
1\,600$\,cm$^2$/Vs. $\rho$(0\,K) presumably stems from
imperfections, and may be related to atomically sharp defects (a
nonvanishing amplitude of the D peak has also been observed in
Raman spectroscopy \cite{emtsev}) while we attribute $\Delta\rho$
to interactions with substrate phonons. Note also that the
mobility plotted against $T$ is remarkably linear.

\begin{figure}
 \includegraphics[width=3in]{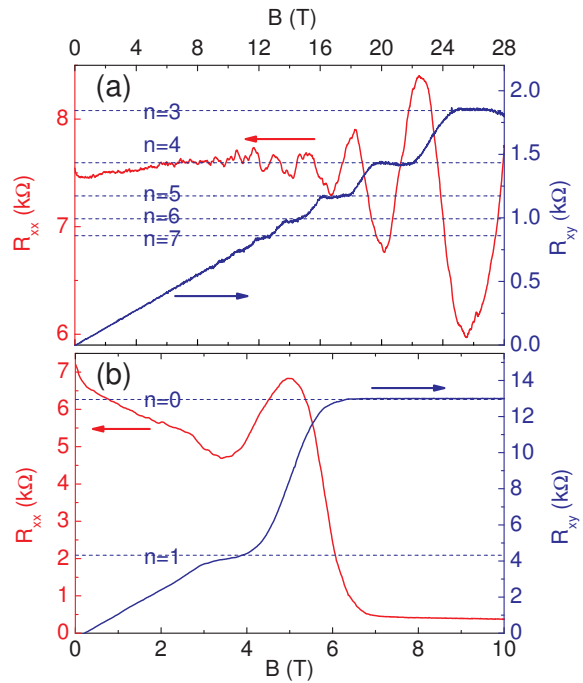}
 \caption{\label{magnetoresistance} (a) Resistance $R_{xx}$ and Hall Resistance $R_{xy}$ at $T= 4.2$\,K from a sample of single-sheet graphene,
 entirely placed on a single substrate step (Fig.~\ref{hall_bar}).
 $R_{xx}$ shows Shubnikov-de Haas oscillations, but no quantum Hall effect.
 The Hall resistance, however, shows step-like plateaus like in the quantum Hall effect.
 Plateau values and the positions of extrema can be identified with the unconventional Landau-level structure of single-layer graphene.
 (b) The same quantities in a sample close to charge neutrality.
 Shubnikov-de Haas oscillations (barely visible below  4\,T) and quantum Hall effect are present and demonstrate the unique single-layer
 graphene properties. Note that due to a lost contact during the experiment, the measurement was carried out as a three-terminal measurement.
 Hence, the quantum Hall resistance is $R_{xx} \approx 300$\,$\Omega$ including the wire resistance, whereas $0$\,$\Omega$ would be expected in a four wire experiment.
 In a further control experiment, we could verify with a slightly degraded sample that the sample indeed showed $R_{xx}\approx 0$\,$\Omega$ at the quantum Hall plateau \cite{additional}.}
 \end{figure}

In a further experiment, we measured the magnetoresistance in
higher magnetic fields up to 28\,T at cryogenic temperatures
($1.4\,$K$ < T < 4.2$\,K). In this regime, electronic degrees of
freedom are condensed in Landau levels, which have in graphene a
significantly different spectrum compared to other materials
\cite{bilayer_QHE}. Figure~\ref{magnetoresistance}(a) shows the
magnetoresistance $R_{xx}$
 and the hall resistance $R_{xy}$as a function of magnetic
 field $B$. With increasing field, the evolution of $R_{xx}$ to
 quantum oscillations can clearly be seen. The quantum Hall regime,
 however, with $R_{xx}=0$ is not yet reached in this sample
 with a charge density $n = 8.9\cdot10^{12}$\,cm$^{-2}$ and $\mu = 2\,300$\,cm$^2$/Vs. $R_{xy}$ already
 displays plateaus, which are precursors of the quantum Hall effect.

\begin{figure}[H]
 \includegraphics[width=3in]{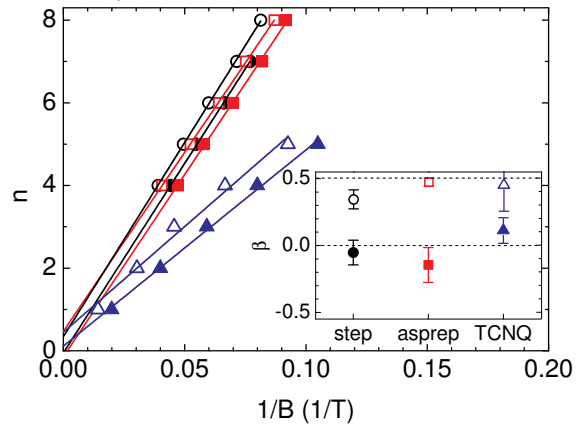}
  \caption{\label{F4TCNQ_QHE} Landau level index $n$ of the SdH maxima (closed symbols) and minima (open symbols) over the inverse magnetic field $1/B$ and linear fits. Circles: as-prepared sample lying on a single substrate terrace. Squares: as-prepared sample covering several substrate steps. Triangles: sample gated close to charge neutrality with F4-TCNQ (plotted against $0.1/B$ for clarity). Inset: The axis intercepts of $\beta=0.5$ and $\beta=0$ for minima and maxima respectively yield a Berry phase of $\pi$ as expected for single-layer graphene. The error bars indicate the standard deviation of the fitting constant.}
 \end{figure}

The values of these plateaus fit in the scheme of
$R_{xy}=h/(4n+2)e^2$ with $n$ being the Landau level index, as
found in exfoliated graphene. When the positions of the associated
maxima (minima) in $R_{xx}$ are plotted against the inverse field
$1/B$ (Fig.~\ref{F4TCNQ_QHE}), a linear dependence can be
recognized. The axis intercept is
$\beta=n(B\rightarrow\infty)=1/2$ ($\beta=0$), as expected and
experimentally confirmed for electrons in exfoliated graphene
(often, this is described as the geometric Berry phase  $\pi$
associated to a closed orbit). Hence, the SdH oscillations in the
raw material (380\,meV above the charge neutrality point) display
graphene physics. Note that the expectations of the Berry phase in
bilayer graphene would be $2\pi$.

\begin{figure}
 \includegraphics[width=3in]{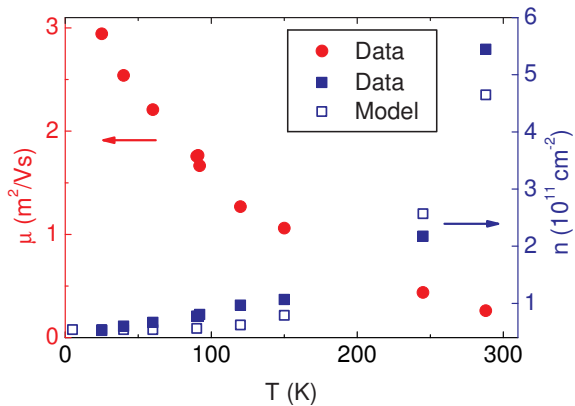}
  \caption{\label{super_mobility}Charge carrier mobility $\mu$ and -density $n$ for a sample close to the charge neutrality point ($E_F$ = 27\,meV). Closed symbols: $\mu$ (red) and $n$ (blue) derived from an evaluation of Hall data, assuming only one charge-carrier type, yields mobilities of 29\,000\,cm$^2$/Vs. The strong temperature dependence comes from the interplay of electrons and holes in the Fermi distribution at finite temperatures. A simple two-band model (open squares) yields a similar $T$ dependence. }
 \end{figure}

The most interesting part of the graphene spectrum is the Dirac
point, or the charge neutrality point, where the density of state
shrinks to zero, and the charge carrier mobility may become huge
\cite{QHE_stormer}. In order to reach this point, we deposited
tetrafluoro-tetracyanoquinodimethane (F4-TCNQ) molecules by
thermal evaporation. Upon contact the strongly electronegative
molecules expel electrons from the graphene and effectively drive
the graphene close to the charge neutrality point \cite{f4tcnq}.
We have chosen a thickness of several monolayers, for which the
sample is reasonable stable. However, some drift in $n$ occurs
within days, accompanied by increasing inhomogeneities
\cite{additional}. Classical Hall effect measurements of the
sample with the lowest $n = 5.4\cdot10^{10}$\,cm$^{-2}$ displayed
excellent mobilities of $\mu = 29\,000$\,cm$^2$/Vs at $T = 25$\,K,
as shown in Fig.~\ref{super_mobility}. Note that for rising
temperatures, not only electrons, but also hole states are
accessible by the Fermi distribution. As a consequence, the
evaluation of the Hall data has to consider two charge carrier
types, so $\mu$ and $n$ can not be unambiguously derived. When
assuming a two-band model ($\mu_{holes} = \mu_{electrons}$ for
simplicity), the resulting simulation describes the temperature
dependence reasonably well. The expectation would be that even
higher mobilities are achievable, if one would come closer to the
Dirac point. Mobilities over $20\,000$\,cm$^2$/Vs are rarely found
for exfoliated graphene on a substrate, and significantly better
values are only found in absence of a substrate \cite{suspended}.
Here, high mobilities are observed although the graphene is in
contact with two surfaces: the SiC substrate with its large
supercell and the virtually disordered F4-TCNQ film on top.

 For the measurements in high magnetic fields, we used another F4-TCNQ covered graphene sample. It was slightly filled with electrons ($n =
4.9*10^{11}$\,cm$^{-2}$, $\mu = 4\,900$\,cm$^2$/Vs at $T =
4.2$\,K). Figure~\ref{magnetoresistance}(b) shows the
magnetoresistance $R_{xx}$ and the Hall resistance $R_{xy}$. In
the low field regime, Shubnikov-de Haas oscillations occur.
Compared to Fig.~\ref{magnetoresistance}(a), the quantum
oscillations are rather compressed as a consequence of the low
charge carrier density. For magnetic fields larger than 7\,T, the
resistance is effectively zero, whereas the Hall resistance has
the value $h/2 e^2$. This is the last plateau of the QHE. From
this single plateau, it can be derived that the QHE is different
from bilayer graphene, where it should be $h/4e^2$. Further
information about the Landau level spectrum can be gained by
analyzing the SdH oscillations similar to the above procedure: a
SdH phase is found which corresponds to
a Berry phase of $\pi$ (Fig.~\ref{F4TCNQ_QHE}). %
It is remarkable that all samples show quantum oscillations
typical for single-layer graphene, although the F4-TCNQ covered
Hall bar as well as one as-prepared sample did not lie on a single
substrate terrace and therefore parts of the sample are bilayers,
which extend as small stripes along the step edges.

We conclude that the raw graphene material, which is strongly
electron filled has a charge carrier mobility around
900\,cm$^2$/Vs at room temperature and 2\,000\,cm$^2$/Vs at low
temperatures. This limitation comes partly from electron-phonon
interaction with substrate phonons, partly from crystal
imperfections. The mobility is unsensitive to substrate steps.
Shubnikov-de Haas oscillations and plateaus in the Hall resistance
indicate that at $B = 28$\,T the quantum Hall regime is not yet
fully reached, but the Landau-level spectrum is (single-sheet)
graphene-like.

When gating close to the Dirac point, high mobilities of 29\,000\,cm$^2$/Vs are observed. The quantum oscillations in high magnetic
fields reveal the Landau-level spectrum of single-sheet graphene,
and the quantum Hall effect is observed. Hence, epitaxial graphene
reproduces the unique features observed in exfoliated graphene,
but is certainly a system which allows for more systematic
development of graphene devices, with rich perspectives for
science and technology.

We gratefully acknowledge support by the DFG under contract SE
1087/5-1, WE 4542-5-1, and within the Cluster of Excellence {\it
Engineering of Advanced Materials} (www.eam.uni-erlangen.de). This
work has been partially supported through EUROMAGNET II under SP7
of the EU; contract number 228043

\bibliography{weber_transport_prl}

\begin{thebibliography}{22}
\bibitem[{add()}]{additional}
\bibinfo{note}{We will provide supporting material}.
\end{thebibliography}

\end{document}